# Assessment of low-carbon tourism development from multi-aspect analysis: A case study of the Yellow River Basin, China

Xiaopeng Si [a], Zi Tang [a] [*]

[a] School of Tourism and Cuisine, Harbin University of Commerce, Harbin, 150028, China

[*] Correspondence should be addressed to Zi Tang. E-mail address: tz09@163.com

**Abstract**

Climate change has become an unavoidable problem in achieving sustainable development. As one of the major industries worldwide, tourism can make a significant contribution to mitigating climate change. The main objective of the paper is to assess the development level of low-carbon tourism from multi-aspect, using the Yellow River Basin as an example. Firstly, this study quantified tourism carbon dioxide emissions and tourism economy, and analyzed their evolution characteristics. The interaction and coordination degree between tourism carbon dioxide emissions and tourism economy were then analyzed using the improved coupling coordination degree model. Finally, this study analyzed the change in total factor productivity of low-carbon tourism by calculating the Malmquist-Luenberger productivity index. The results showed that: (1) The tourism industry in the Yellow River Basin has the characteristics of the initial environmental Kuznets curve. (2) There was a strong interaction between tourism carbon dioxide emissions and tourism economy, which was manifested as mutual promotion. (3) The total factor productivity of low-carbon tourism was increasing. Based on the above results, it could be concluded that the development level of low-carbon tourism in the Yellow River Basin has been continuously improved from 2000 to 2019, but it is still in the early development stage with the continuous growth of carbon dioxide emissions.

# 1 Introduction

Environmental issues represented by climate change have emerged as a global concern. The environmental impact of tourism is mainly manifested in the consumption of resources and the greenhouse gas emissions, which are a well-known cause of climate change [1][2]. However, research on the environmental impacts caused by tourism is usually not independent, but connected to the growth of tourism economy (TE). In other words, the ultimate goal of considering the environmental impact of tourism is to make it a sustainable industry.

Research on sustainability has expanded to several frontiers [3]. Sustainable de-growth is regarded in some circles as a better way of achieving personal and social welfare on a worldwide scale, so it is not only necessary to reconsider the role of tourism (as an inherently economical industry with considerable resources effects) in development in general, but also to seek means to reduce the impact of tourism on the environment [4][5]. Climate change is one of the issues that sustainable development has to consider. The economic impact of climate change is likely to be limited in the 21st century and may even be beneficial in the short to medium run, but the adverse effects of climate change could exceed the beneficial effects in the long term [6]. The increase of the concentration of carbon dioxide in the atmosphere is considered to be the main factor of climate change, although there is some controversy [7][8][9][10]. For this reason, some studies on climate change and economic development are conducted from the perspective of carbon emissions. Mardani et al. reviewed 175 articles on carbon dioxide and economic growth from 1995 to 2017 and found that the bidirectional relationship of economy development and carbon dioxide emissions does exist [11].

There have been many studies on the relationship between tourism and climate change. Through a bibliometric analysis of 1290 articles, Scott and Gössling found that the literature related to climate change has increased rapidly and climate change has regional implications for tourism, through varying effects on natural and cultural heritage and shifts in patterns of demand [12]. Evidence from Pintassilgo et al. shows that climate change is likely to cause substantial negative economic impacts in the Portuguese tourism sector, specifically, inbound tourism arrivals will decrease by 2.5% to 5.2%, which is expected to reduce Portuguese GDP by 0.19% to 0.40% [13]. The research results of Yañez et al. on Coachella Valley show that with predicted climate change, the season of comfortable will inevitably shorten, which may have a significant socio-economic impact on regions with strong winter tourism industries [14]. Similarly, the issue of carbon emissions from the development of tourism has also drawn focus. After 2010, the number of publications published on tourism and carbon emissions increased steadily each year [15]. Lenzen et al. found that due to the fact that tourism is substantially more carbon dense than other potential sectors for growth in the economy, pursuing the goal of economic expansion by its swift growth would come with a sizable carbon load [16]. Evidence from Dogru et al. indicates while both the entire economy and the tourism industry are similarly resilient to climate change, the tourism industry is more susceptible to it [17]. After that, Dogru et al. found that tourism development has significantly different impacts on carbon dioxide emissions in different OECD countries [18]. When it is found that tourism is not a low-carbon industry, research on low-carbon tourism has gradually received attention.

Low-carbon tourism is an essential tool for sustainable development, particularly in areas with a sizeable tourism industry. Although there are many research methods that can be used to study low-carbon tourism, little work has so far been undertaken to attempt a combination of methods that might find some more comprehensive and credible results. In addition, the existing assessment of low-carbon tourism focuses on the measurement of development level [19][20][21]. The method of measuring the development quality of low-carbon tourism by development level is intuitive and effective, but it is also incomplete. This study argues that the analysis of other performance indicators besides the development level is also of great significance to assess the quality of low-carbon tourism development. Tourism carbon dioxide emissions (TCDE) and TE are the main components of low-carbon tourism, and their interaction affects their whole performance. Therefore, it is necessary to analyze the interaction and overall performance of these two systems when assessing the low-carbon tourism development. The paper aims to achieve this goal through the coupling coordination degree (CCD) model and the Malmquist-Luenberger productivity index (MLPI).

Coupling, which originated in physics, is referred to as the interaction between several systems, and the coupling degree (CD) and CCD reflects the process of the overall evolution of the system [22]. The CCD model has been widely used to study the interaction between tourism systems or between tourism and non-tourism systems [23][24]. The MLPI was proposed on the basis of Malmquist index [25]. Compared with the latter, it is constructed from a directional distance function and takes into account whether the output is a desired positive output or an undesired negative output [26][27]. When coupled coordination model and MLPI are used at the same time, not only the interaction state of TCDE and TE as two systems, but also their overall performance as a system can be studied. However, few studies have linked the two approaches. As the largest developing country, China is committed to achieving carbon peaking and carbon neutrality. Considering that tourism is a strategic industrial pillar for China's economy, it is essential to comprehend the connection between the TE and TCDE to achieve both China's carbon peaking and carbon neutrality goals and the sustainable tourism of the World Tourism Organization. A number of studies have found evidence that China's tourism industry is transitioning to a low-carbon industry [28][29][23][30]. However, there are 34 provincial administrative regions throughout

China, and the development of tourism in each region is obviously different [31][32]. This paper uses the CCD model and MLPI to analyze the development of low-carbon tourism in nine provincial administrative regions within the Yellow River Basin (YRB), which may be more representative of developing countries. In addition, the possible existence of an Environmental Kuznets Curve (EKC) for tourism industry in YRB is discussed. According to the EKC hypothesis, the harm to the environment is increasing while the economic growth rate is rapid during the early stages of development. If the economy develops to a higher level with low growth, the damage to the environment is reduced [33]. China's economy has maintained rapid growth for many years, and the resulting environmental issues have become one of the major challenges troubling the Chinese government. In pursuit of sustainable development, China has set the "dual carbon" goal of carbon peaking and carbon neutrality. With continued economic growth, if this goal is effectively achieved, a relationship between China's economy and its carbon emissions will emerge as the EKC.

## 2 Material and methods

### 2.1 Study area

The Yellow River is the second longest river in China, flowing through nine provincial administrative regions, including Qinghai, Sichuan, Gansu, Ningxia, Inner Mongolia, Shanxi, Shaanxi, Henan and Shandong (Fig. 1). Geographical characteristics are not the focus of this study, so this paper takes these nine provincial administrative regions with more abundant statistical data rather than cities with more accurate geographic information but less statistical data as the study area.

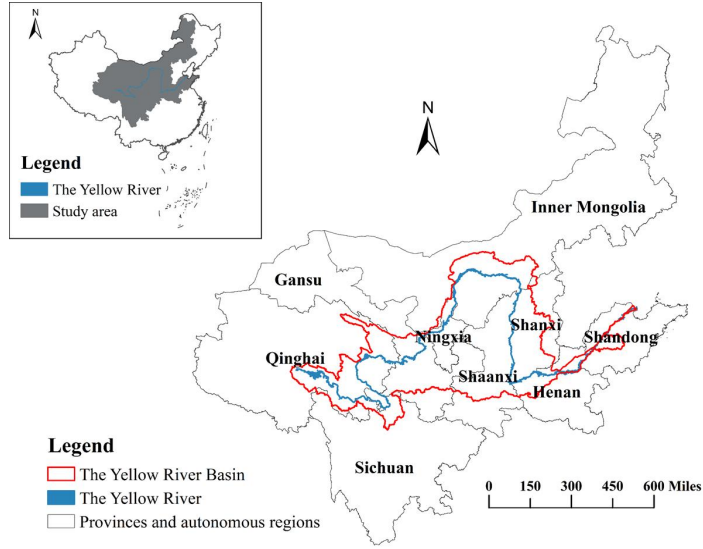

Fig. 1. Overview of study area. This map was generated by the authors using ArcGIS 10.8 (http://www.esri.com/software/arcgis) and does not require any license.

## 2.2 Methodology and procedure

### 2.2.1 Method for estimating tourism-related carbon dioxide emissions

The calculation of TCDE in YRB refers to the methods in previous studies, and the specific steps are listed below [34].

$$Q_T = \sum_{i=1}^{n} \alpha_i \cdot N_i \cdot D_i \cdot P_{ci} \qquad (1)$$

$$Q_H = 365 \cdot Y \cdot R \cdot P_e \cdot P_{cv} \cdot 10^{-3} \cdot \frac{44}{12} \qquad (2)$$

$$Q_A = \sum_{j=1}^{m} M \cdot \omega_j \cdot P_{cj} \qquad (3)$$

$$Q = Q_T + Q_H + Q_A \qquad (4)$$

where $Q_T$ is the TCDE from transportation; $\alpha_i$ represents the percentage of travelers as passengers in the $i$th mode of transportation. $N_i$ is the $i$th mode of transportation's passenger count; $D_i$ is the $i$th mode of transportation's journey distance; $P_{ci}$ is the $i$th mode of transportation's carbon dioxide emissions coefficient; $n$ is the total number of transportation modes, including plane, car, train and water transportation; $Q_H$ is the TCDE from accommodation; Y is the total quantity of beds, which represents the reception capacity of tourism accommodation industry; R stands for the annual bed occupancy rate; $P_e$ is the accommodation energy consumption coefficient; $P_{cv}$ is amount of carbon per energy equivalence unit; 44/12 is the carbon to carbon dioxide conversion coefficient; $Q_A$ is the TCDE from tourism activities; M represents the amount of tourists; $\omega_j$ is the percentage of the $j$th tourism activity; $P_{cj}$ is the $j$th mode of activity's carbon dioxide emissions coefficient; $m$ is the total number of tourism activity types, including sightseeing, leisure vacation, business trip, visiting relatives and friends, and others; Q is the total TCDE. The TCDE index is obtained by using the same method as TE index to deal with TCDE data.

### 2.2.2 Assessment and quantification of tourism economy

In this study, tourist arrivals, tourism revenues and tourism practitioners, which are highly related to TE, are selected as indicators to assess the level of TE in YRB [35][36][37]. Tourist arrivals and tourism revenues are calculated using data from both domestic and international sources.

Entropy method is applied for determining the weight of different dimension indicators through information entropy, which is suitable for the calculation of TE indexes [38][39][40]. In addition, considering that the differences of TE level in different regions is not within the scope of this study, so the use of entropy method to measure the various regional TE index is carried out separately. The calculation steps are listed below.

Step 1: The Min-Max Normalization is used to standardize the indicators to eliminate the difficulties caused by different orders of magnitude and dimensions. The calculation process is shown in formula (5). If the attribute of $X_{ij}$ is positive, then

$$X'_{ij} = \frac{X_{ij} - min\{X_j\}}{max\{X_j\} - min\{X_j\}} + \alpha \tag{5}$$

where $X_{ij}$ is the $j$th indicator's value in $i$th year; $X'_{ij}$ represents the value of $X_{ij}$ after standardization; α is an appropriate positive number that serves to avoid $X'_{ij}$ equal to zero that would make subsequent calculations impossible ( α = 0.00001 in this study); $max\{X_j\}$ and $min\{X_j\}$ represent the maximum and minimum value of the $j$th indicator respectively.

Step 2: Calculate the weights of each indicator.

$$R_{ij} = X'_{ij} \bigg/ \sum_{i=1}^{n} X'_{ij} \tag{6}$$

$$e_j = -\frac{1}{\ln n} \sum_{i=1}^{n} R_{ij} \ln R_{ij} \tag{7}$$

$$w_j = (1 - e_j) \bigg/ \sum_{j=1}^{m} (1 - e_j) \tag{8}$$

where $R_{ij}$ is the contribution degree of $X'_{ij}$; $n$ is the year span; $e_j$ represents the entropy of $j$th indicator; $w_j$ represents the weight of $j$th indicator; $m$ represents the number of indicators contained in each assessment index.

Step 3: Calculate TE index $E_i$. The comprehensive assessment index is calculated as shown in equation (9). It is usually calculated using the linear weighting method.

$$E_i = \sum_{j=1}^{m} w_j X'_{ij} \tag{9}$$

where $E_i$ is TE index in year $i$. However, the value range of $E_i$ calculated by Min-Max Normalization and Entropy method is [0,1], and the determined extreme value may cause great deviation to the outcomes of the subsequent CCD. In addition, a lower-level system should correspond to a lower assessment index, but this assessment index should not be 0 unless the system stops operating. The $E_i$ in this study cannot be equal to 0 or 1 because of the existence of α, but if all values of $X_j$ in year $i'$ are maximum or minimum, the comprehensive assessment index in year $i$ ($E_{i'}$) must be α or 1 + α. In this case, the CCD of the endpoint values may also show a deviation that is difficult to ignore. In order to solve this problem, two new variables $X_{pj}$ and $X_{qj}$ are introduced into the data in this study to improve the measurement of $E_i$. Let the data matrix with new variables is $X^0{}_{ij}$. If $X_{pj}$ is the maximum and $X_{qj}$ is the minimum in $X^0{}_{ij}$, then the range of standardized values corresponding to $X_{ij}$ is (0,1) instead of [0,1] after using Min-Max Normalization for $X^0{}_{ij}$. Considering the purpose of processing, the rationality of results and the characteristics of Min-Max Normalization, the value of $X_{pj}$ should be determined by the original dataset $X_{ij}$. In this paper, let $X_{pj} = max\{X_j\} + min\{X_j\}$. As for $X_{qj}$, because its standardization value is 0, so make it equal to zero is reasonable. Assuming that the normalized result of $X_{ij}$ obtained by standardizing $X^0{}_{ij}$ is $X''{}_{ij}$, the calculation formula of $X''{}_{ij}$ is shown in equation (10). The improved calculation method of $E_i$ is shown in equation (11).

$$X''{}_{ij} = \frac{X_{ij} - min\{X^0{}_j\}}{max\{X^0{}_j\} - min\{X^0{}_j\}} = \frac{X_{ij} - X_{qj}}{X_{pj} - X_{qj}} = \frac{X_{ij}}{max\{X_j\} + min\{X_j\}} \tag{10}$$

$$E_i = \sum_{j=1}^{m} w_j X''{}_{ij} \tag{11}$$

### 2.2.3 Calculation of coupling coordination degree

CCD model is simple for calculation and the results intuitively understandable, which is why it is extensively applied for CCD measurement between systems [39][23][41][42]. And with the deepening of related research, the discussion on the improvement of CCD model has been going on [43][44][45][46]. Referring to the method of Shen et al. and Wang et al., the calculation steps are listed below [43][44].

$$C = \frac{\sqrt{\left[1 - \sqrt{(U_1 - U_2)^2}\right] \times U_1 U_2}}{\max(U_1, U_2)} \quad (12)$$

$$\alpha = \frac{U_2}{U_1 + U_2} \quad (13)$$

$$\beta = \frac{U_1}{U_1 + U_2} \quad (14)$$

$$T = \alpha U_1 + \beta U_2 \quad (15)$$

$$D = \sqrt{C \times T} \quad (16)$$

where $C$ is the CD of TE system and TCDE system; $U_1$ is the TE index; $U_2$ is the TCDE index; $T$ represents the two systems' overall development level; $\alpha$ and $\beta$ represent the contributions of U1 and U2 to $T$, respectively; $D$ is the CCD of the two systems. In order to find out more detailed information, a more meticulous classification criterion as shown in Table 1 is used for the CCD [23][40].

Table 1 The division standard for coupling coordination degree

| Condition | CCD | Levels | Serial Number | Condition | CCD | Levels | Serial Number |
|---|---|---|---|---|---|---|---|
| | [0, 0.1) | Extreme | Level 1 | | [0.5, 0.6) | Barely | Level 6 |
| | [0.1, 0.2) | Severe | Level 2 | | [0.6, 0.7) | Primary | Level 7 |
| Imbalance | [0.2, 0.3) | Moderate | Level 3 | Coordination | [0.7, 0.8) | Mediocre | Level 8 |
| | [0.3, 0.4) | Slight | Level 4 | | [0.8, 0.9) | Good | Level 9 |
| | [0.4, 0.5) | Imminent | Level 5 | | [0.9, 1.0] | Super | Level 10 |

## 2.2.4 Malmquist-Luenberger productivity index

The existence of non-desired outputs is one of the keys affecting the applicability of efficiency assessment methods [47][48]. Since MLPI was introduced by Chung et al., it has been widely used to measure total factor productivity with undesirable outputs [25][26][48][49]. The two components of the MLPI are used to describe changes of technical efficiency (MLTE) and technical change improvements (MLTC). The calculation results of MLPI may have infeasibility problem [50][51]. When there are few feasible solutions that do not affect the overall results, this paper uses geometric mean completion. Álvarez et al. developed a data envelopment analysis toolbox for MATLAB, which contains a package for calculating MLPI [52]. This paper uses this toolbox that has been verified by other studies to calculate MLPI, MLTE and MLTC [53][54]. The three input indicators are tourism practitioners, tourism fixed assets investment and energy consumption of tourism. The two positive output indicators are tourist arrivals and tourism revenues, and the undesirable output indicator is TCDE.

## 2.3 Data sources

The data used to calculate TCDE and TE come from China Statistical Yearbook, China Energy Statistical Yearbook, the Yearbook of China Tourism Statistics, China Cultural Heritage and Tourism Statistical Yearbook, and statistical yearbooks of provincial administrative regions. Most of the data used to calculate the total factor productivity of tourism also come from the above statistics. Tourism practitioners refer to the number of employees in travel agencies. The data selected for fixed asset investment in tourism are fixed asset investment in transit, wholesale and retail, and lodging and restaurant related to tourism. Relevant data for it come from Statistical Yearbook of the Chinese Investment in Fixed Assets. Data for number of domestic tourists in 2020 and 2021 are sourced from provincial statistical yearbooks and bulletins. And this data in Henan Province is missing and replaced by the total number of tourists. The energy consumption data of wholesale and retail trades, hotels and catering services are also from China Energy Statistical Yearbook. Some missing data in the yearbook were supplemented from the National Economic and Social Development bulletin.

# 3 Results

## 3.1 Development characteristics of tourism carbon dioxide emissions and tourism economy

The TCDE level of nine regions in YRB increased from 2000 to 2019, and the trend of change was consistent. For example, the trend of TCDE in most regions stopped increasing significantly between 2008 and 2010, but increased again in 2011. This may be caused by the effects of the 2008 financial crisis. And in 2019, the TCDE level of six regions reached more than 0.8, and the remaining three also reached more than 0.6 (Fig. 2).

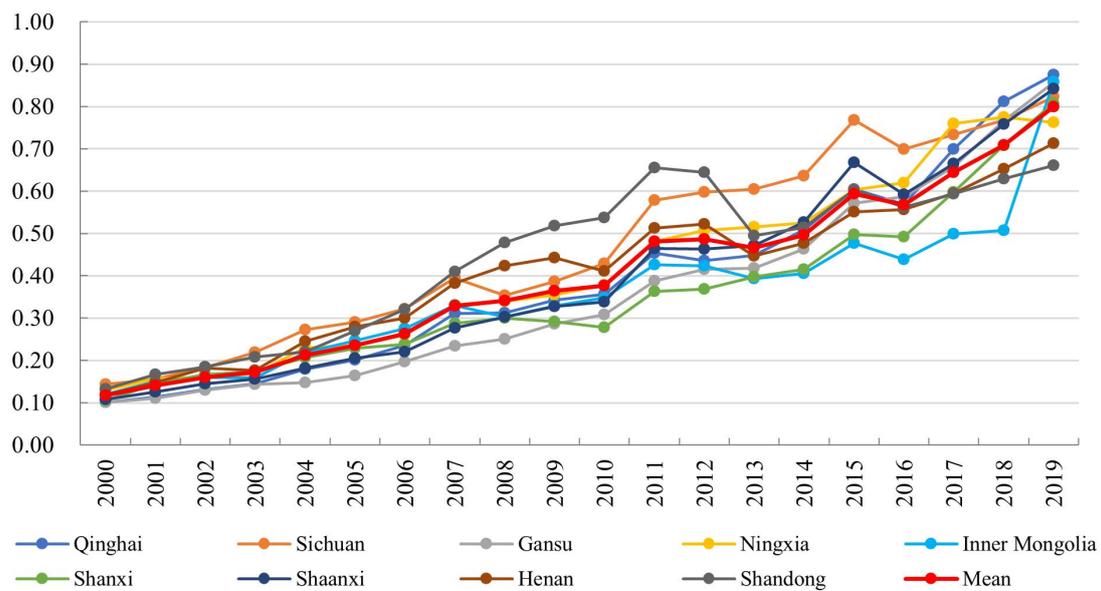

Fig. 2. TCDE level of 9 regions in YRB

The results of the TE index showed that the TE level in all regions of YRB increased steadily from 2000 to 2019, and reached above 0.8 in 2019. Most of the increasing amount occurred after 2010. In 2010, except for Shandong Province, the TE level of the other eight regions in YRB was lower than 0.3. And since 2011, the TE increment of the nine regions has increased year by year.

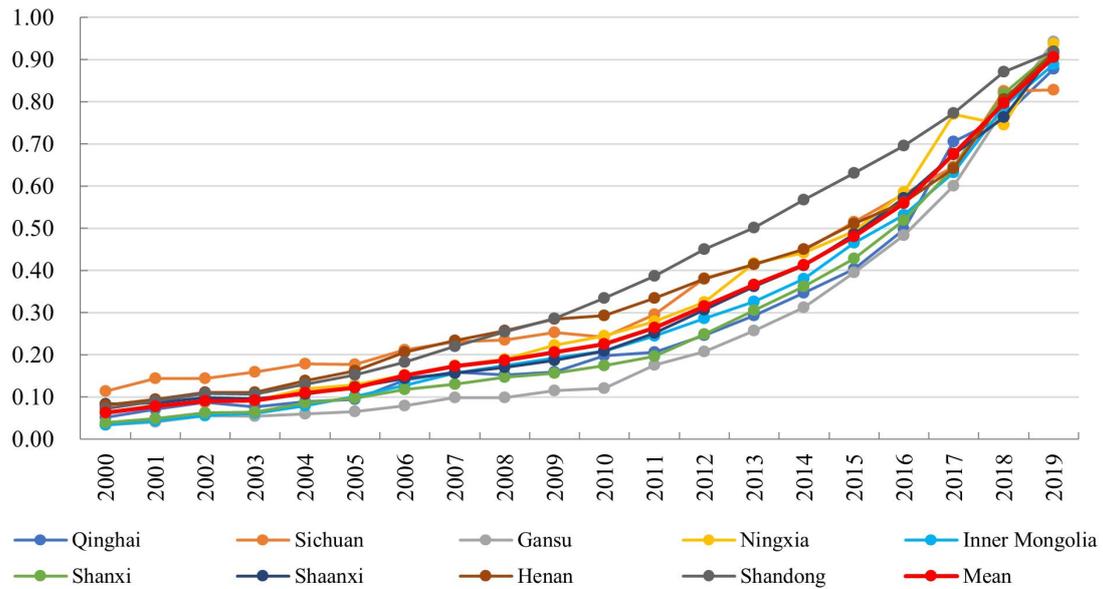

Fig. 3. TE level of 9 regions in YRB

Comparing Fig. 2 and Fig. 3, the change process of the performance level of the two subsystems has both similarities and differences. The main similarities were that both TCDE and TE maintained an increasing trend, and the nine provinces in YRB showed a strong consistency. The main difference is that the annual increment of TCDE is stable, which means that the relationship between time and TCDE is close to direct proportion. The annual increment of the TE is increasing, showing a stronger growth trend than TCDE. These can be obtained by comparing the trend of Mean in Fig. 2 and Fig. 3.

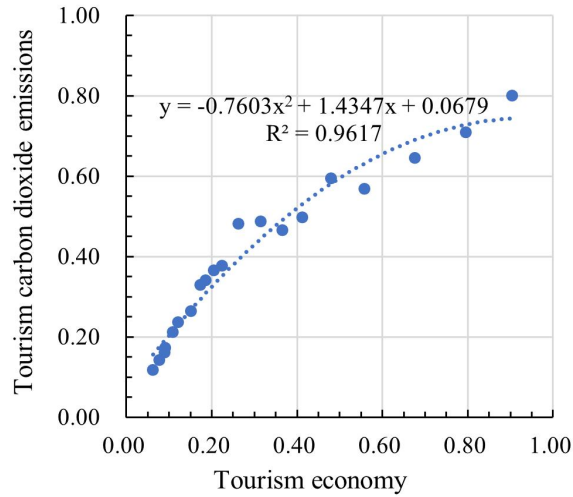

Fig. 4. The EKC of low-carbon tourism in YRB

YRB exhibits the characteristics of EKC in its relationship between TCDE and TE (Fig. 4). Until 2019, TCDE in YRB are still increasing in a relatively stable annual increment, while the TE is increasing in an increasing annual increment. Based on the EKC, it can be hypothesized that tourism carbon emissions would decline gradually when the level of the tourism economy is able to reach a certain tipping point. In the context of the carbon peaking and carbon neutrality goals, the Chinese government is pursuing high-quality development and sustainable development, which contributes to the emergence of the tipping point.

## 3.2 Coupling coordination characteristics of tourism carbon dioxide emissions and tourism economy

The CD between TCDE and TE in YRB changed from 0.70 in 2000 to 0.89 in 2019, showing an increasing trend (Fig. 5). According to the characteristics of time change, it can be divided into three developments: (1) From 2000 to 2011, the CD always fluctuated slightly around 0.7. Although there is a significant gap in the CD of different provincial administrative regions, the gap is decreasing, and the range of the gap is reduced from 0.37 in 2000 to 0.15 in 2011. (2) From 2012 to 2016, the CD increased significantly. The CD in 2011 was 0.65, which was the minimum value during the study period. And the CD in 2016 was 0.91, only 0.01 smaller than the maximum. Therefore, the period from 2012 to 2016 is an important period of increasing interaction between TCDE and TE in YRB. (3) From 2017 to 2019, the CD remained relatively stable at around 0.9. After the second stage of growth, the TCDE and TE in YRB have maintained a high coupling level.

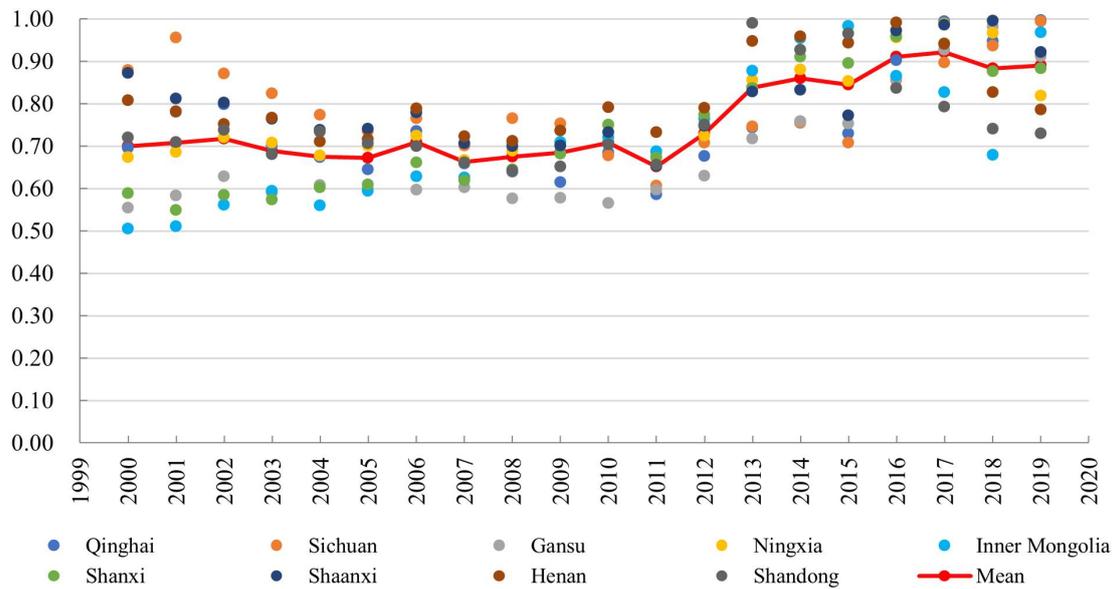

Fig. 5. CD of 9 regions in YRB

The CCD of TCDE and TE in YRB is shown in Fig. 6. It is obvious that the CCD of the two subsystems in nine regions increased steadily from 2000 to 2019. The direct reason is the TCDE and TE continued to grow during the study period. The three regions of Gansu, Inner Mongolia, and Shanxi had the worst coupling coordination status overall in 2000, with a severe level of incoordination. Sichuan had the best coupling coordination condition and was in a slight incoordination level. And other regions were in a moderate incoordination level. In 2019, Henan and Shandong had the worst coupling coordination status, with a mediocre coordination level. Shanxi, Shaanxi and Ningxia were better and in good coordination level. And other four regions were in a super coordination level. Through comparison, it can be found that Inner Mongolia and Gansu changed from the region with the worst coupling coordination condition to the region with the best coupling coordination condition. Henan and Shandong had relatively good coupling coordination condition in 2000, but these two provinces became the regions with the worst coupling coordination condition in 2019.

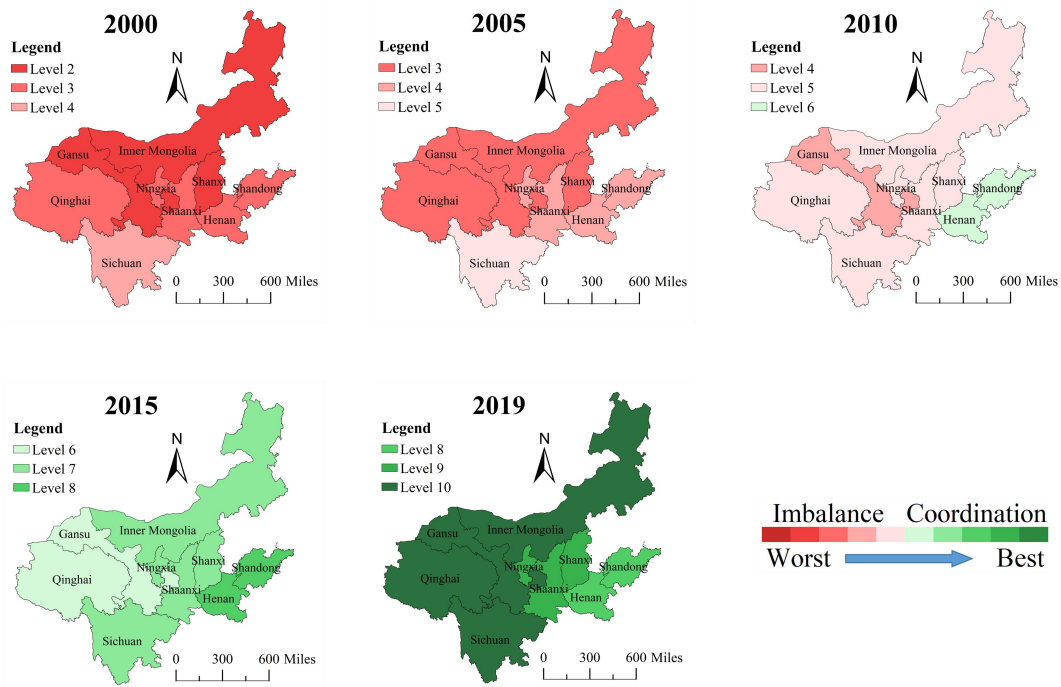

Fig. 6. CCD of 9 regions in YRB. This map was generated by the authors using ArcGIS 10.8 (http://www.esri.com/software/arcgis) and does not require any license.

## 3.3 Total factor productivity development characteristics of low carbon tourism

The MLPI data shown in Table 2 reveal the evolution of total factor productivity of low-carbon tourism in YRB. The majority of the years during the study period had MLPI values greater than 1, suggesting the total factor productivity of low-carbon tourism in YRB was continuously optimized. The average MLPI is 1.07, which means that the total factor productivity of low-carbon tourism in YRB is growing at an average annual rate of 7%. From the perspective of spatial characteristics, the value of MLPI in 9 regions is different, but the change trend is similar. For example, from 2002 to 2003, the MLPIs of nine regions were all less than 1, and most of them were the minimum values in the whole study period (Table 2). Of them, the MLPI of Shanxi is 0.85, which is the lowest among the nine regions. An interesting phenomenon is that the largest average MLPI also appeared in Shanxi, indicating that the total factor productivity of low-carbon tourism in Shanxi increases by 16% per year on average, which is the fastest among the nine regions in YRB. Qinghai saw the slowest growth in low-carbon tourism total factor productivity, with an average annual increase of only 1%, 15% lower than Shanxi.

Table 2 MLPI of low-carbon tourism in YRB

| Regions | 2000~2001 | 2001~2002 | 2002~2003 | 2003~2004 | 2004~2005 | 2005~2006 | 2006~2007 | 2007~2008 | 2008~2009 | 2009~2010 |
|---|---|---|---|---|---|---|---|---|---|---|
| Qinghai | 1.031 | 1.001 | 0.952 | 1.034 | 1.045 | 1.041 | 1.014 | 0.905 | 1.049 | 1.003 |
| Sichuan | 0.994 | 1.159 | 0.934* | 1.088* | 1.091* | 1.059* | 1.073* | 1.015* | 1.081* | 1.072* |
| Gansu | 1.014 | 1.033 | 0.943 | 1.011 | 1.028 | 1.026 | 1.085 | 0.959 | 1.103 | 0.993 |
| Ningxia | 1.009 | 1.007 | 0.972 | 1.043 | 1.058 | 1.039 | 1.056 | 0.998 | 1.012 | 1.008 |
| Inner Mongolia | 1.040 | 1.061 | 0.976 | 1.215 | 1.142 | 1.075 | 1.134 | 1.152 | 1.249 | 1.184 |
| Shanxi | 0.959 | 1.042 | 0.853 | 1.169 | 1.119 | 1.126 | 1.104 | 1.011 | 0.979 | 1.133 |
| Shaanxi | 1.006 | 0.990 | 0.933 | 1.217 | 1.133 | 1.027 | 1.050 | 1.021 | 1.114 | 1.126 |
| Henan | 0.896 | 0.940 | 0.934* | 1.088* | 1.112 | 1.085 | 1.073* | 1.079 | 1.081* | 1.072* |
| Shandong | 0.992* | 1.067 | 0.918 | 0.957 | 1.097 | 1.054 | 1.073* | 1.015* | 1.081* | 1.072* |
| YRB | 0.993 | 1.033 | 0.935 | 1.091 | 1.092 | 1.059 | 1.074 | 1.017 | 1.083 | 1.073 |

| Regions | 2010~2011 | 2011~2012 | 2012~2013 | 2013~2014 | 2014~2015 | 2015~2016 | 2016~2017 | 2017~2018 | 2018~2019 | Average |
|---|---|---|---|---|---|---|---|---|---|---|
| Qinghai | 1.041 | 1.002 | 1.041 | 1.010 | 1.017 | 1.039 | 1.008 | 1.013 | 1.022 | 1.013 |
| Sichuan | 1.065* | 1.092* | 1.142* | 1.097* | 1.042* | 1.093* | 1.146* | 1.047* | 1.082* | 1.071 |
| Gansu | 1.048 | 1.078 | 1.083 | 1.130 | 0.979 | 1.056 | 1.140 | 0.983 | 1.225 | 1.046 |
| Ningxia | 1.018 | 1.013 | 1.076 | 0.970 | 1.002 | 1.015 | 1.024 | 1.075 | 0.982 | 1.019 |
| Inner Mongolia | 1.075 | 1.130 | 1.245 | 1.188 | 1.082 | 1.112 | 1.115 | 1.115 | 0.914 | 1.113 |
| Shanxi | 1.156 | 1.204 | 1.242 | 1.234 | 1.120 | 1.182 | 1.849 | 1.023 | 1.517 | 1.143 |
| Shaanxi | 1.060 | 1.141 | 1.079 | 1.054 | 1.012 | 1.109 | 1.066 | 1.408 | 0.979 | 1.076 |
| Henan | 1.065* | 1.092* | 1.142* | 1.114 | 1.075 | 1.185 | 1.103 | 0.814 | 1.083 | 1.050 |
| Shandong | 1.065* | 1.092* | 1.248 | 1.101 | 1.055 | 1.056 | 1.044 | 1.035 | 1.042 | 1.054 |
| YRB | 1.066 | 1.094 | 1.144 | 1.100 | 1.043 | 1.094 | 1.166 | 1.057 | 1.094 | 1.013 |

* indicates that the infeasible value is replaced by the geometric mean of the data in other regions in the same year. The values of YRB are expressed as the arithmetic mean of 9 regions, and the average MLPI of each region is measured by the geometric mean.

Fig. 7 (a) shows the geometric mean of MLPI, MLTE and MLTC, which reflect the annual changes of total factor productivity, technical efficiency and technical progress of low-carbon tourism in nine regions, and Fig. 7 (b) shows the time evolution characteristics of MLPI, MLTE and MLTC in YRB. The results of Fig. 7 (a) show that while there are significant regional differences in low-carbon tourism technical progress in YRB, there are no discernible regional disparities in the technical efficiency. The decrease of MLTC in Fig. 7 (b) from 2002 to 2003 can explain why MLPI was at a low level in the same period. MLTC is frequently more than 1, demonstrating that low-carbon tourism industry of YRB has achieved significant technical advancements. Furthermore, given that MLTC is typically higher than MLTE, it may be concluded that technical advancement has a more beneficial effect on the growth of total factor productivity than does technical efficiency. The fact that MLTC and MLPI have more comparable spatial distribution features lends further weight to this conclusion (Fig. 7(a) & Fig. 8).

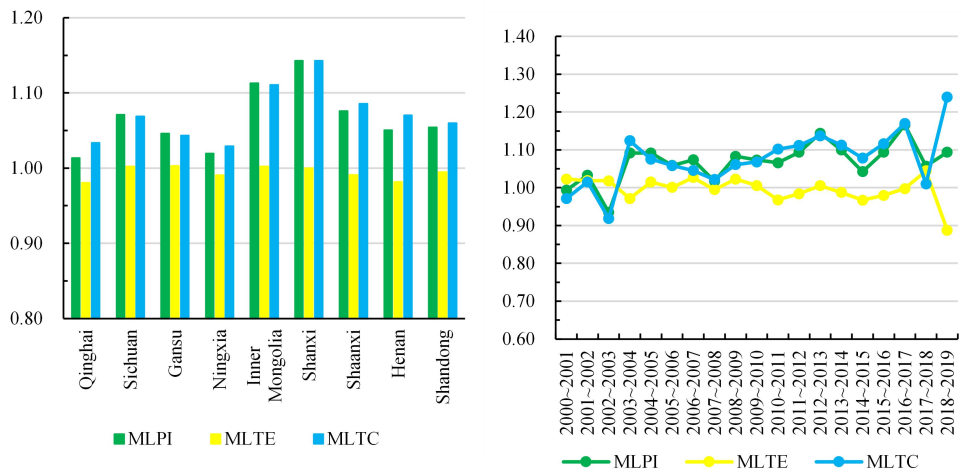

(a) The geometric mean of MLPI, MLTE, and MLTC in 9 regions     (b) MLTE, MLTC and MLPI in YRB

Fig. 7. Temporal and spatial characteristics of MLPI, MLTE and MLTC

The nine provincial administrative regions in YRB were ranked from high to low based on the average values of MLPI, MLTC and MLTE, and the findings are displayed in Fig. 8. Geographical variables may have an impact on changes in total factor productivity, as evidenced by the fact that the difference in the ranking of locations with minor geographic gaps is similarly modest. In addition, it may be seen that there is no evident relationship between technical progress and technical efficiency by contrasting MLTC and MLTE in various places. For example, the growth rate of technical efficiency in Gansu is the fastest among the nine regions in YRB, but its technological progress rate is only ranked seventh. Although the growth rate of technical efficiency is mediocre, Shanxi has the fastest speed of technical progress, which makes its total factor productivity growth rate the fastest. Qinghai has the lowest total factor productivity growth performance in YRB because of poor performance of its technical efficiency and technical progress.

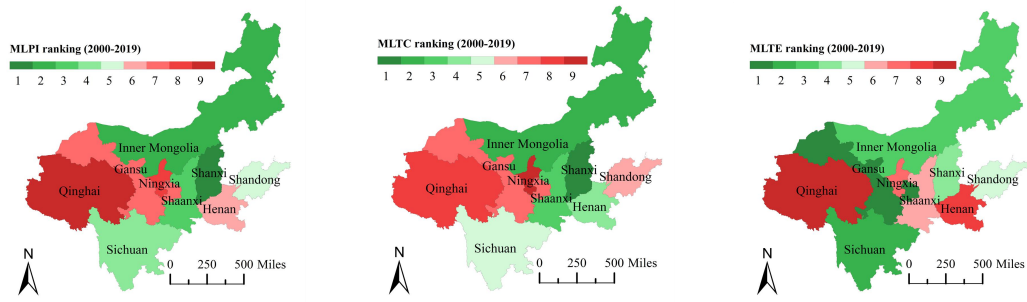

Fig. 8. Ranking the average values of MLPI, MLTC and MLTE in nine regions. This map was generated by the authors using ArcGIS 10.8 (http://www.esri.com/software/arcgis) and does not require any license.

## 3.4 Tourism industry during the COVID-19 pandemic period

COVID-19 has hit the global tourism industry severely, and the negative effects of it on tourism are likely to be long-lasting [55]. Figure 9 presents data on the number of domestic tourists in YRB from 2011 to 2021. Before the outbreak of COVID-19, the tourism industry in YRB achieved satisfactory growth. However, affected by COVID-19, its market size declined to the level of four or five years ago in 2020. Despite showing an upturn in 2021, it doesn't come close to making up for the effects of COVID-19. Therefore, COVID-19 has undoubtedly caused serious and irreversible damage to the tourism industry in YRB.

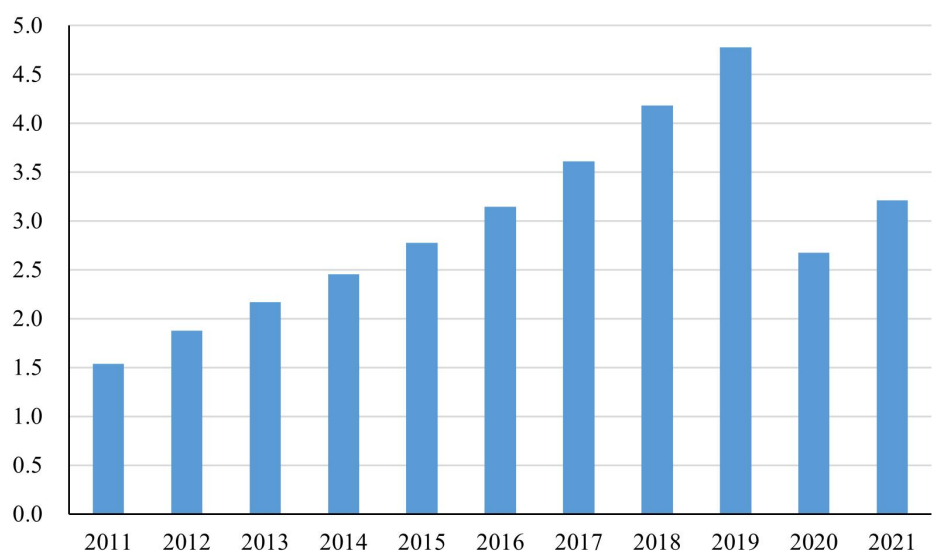

Fig. 9. Number of domestic tourists in YRB (billion person-time)

Although the development of the tourism economy in YRB has been hampered by COVID-19, the low-carbon development of tourism may still be progressing. Figure 10 shows the changes in the consumption of various types of energy by wholesale and retail trades, hotels and catering services (closely related to the tourism industry) in YRB. To achieve its "dual-carbon" goal, China is gradually reducing its reliance on coal consumption in some industries. While coal consumption is still one of the main sources of electricity in China, the development of cleaner power generation technologies, such as photovoltaics, will contribute more effectively to low-carbon development.

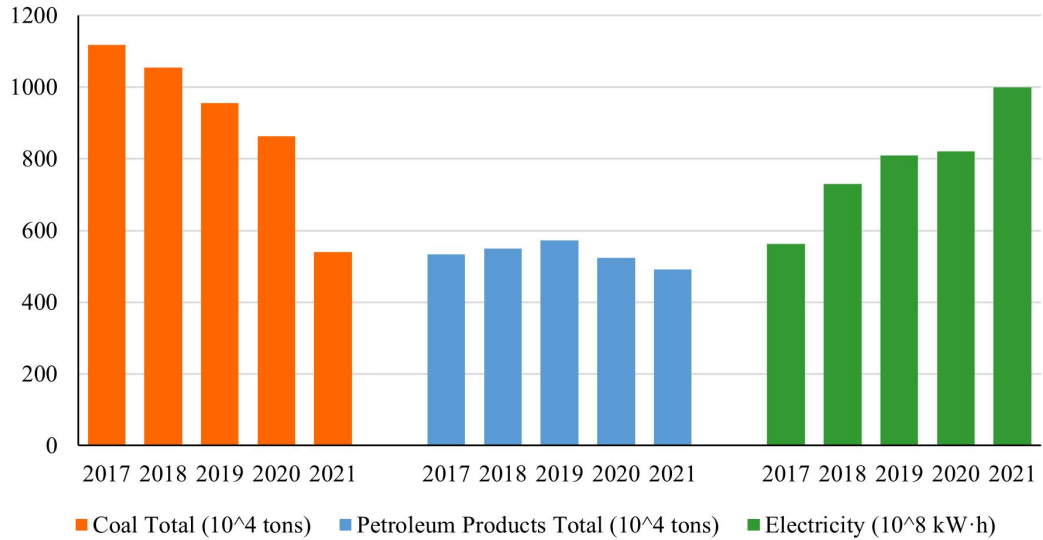

Fig. 10. Energy consumption by wholesale and retail trades, hotels and catering services in YRB

## 4 Discussion

The study found that the tourism industry in YRB showed some characteristics of EKC. The results are in line with Chan and Wong, who observed that although the average person's carbon dioxide inventory decreases due to the growth of tourism, China's provinces are still all at left of the EKC in 2015 [56]. According to the EKC hypothesis, with more and more attention paid to environmental protection, TCDE in YRB may stop growing or even decrease, and the TE can still achieve a significant increase at a lower growth rate. However, it is obvious that the tourism industry in YRB is still at the early stages of development.

The increasing CCD of TCDE and TE also proves the above view. The results of the CCD model show that there is a strong interaction between TCDE and TE in YRB, which may mean that the second stage of the EKC will not appear in the tourism industry in YRB at present. From 2000 to 2019, the tourism industry in YRB developed as rapidly as China 's economy. Under the high coupling level, the CCD of TCDE and TE is steadily enhanced. The research on decoupling analysis of TCDE and TE also draws similar conclusions. Xiong et al. discovered that there is a feeble decoupling linkage between the regional TCDE and the economic progress, which is primarily driven by tourism in China [57]. As the pursuit of high-quality development of TE continues, the growth rate of TE is likely to remain higher than the growth rate of TCDE, leading to an even wider gap in their respective growth rates [58]. When the second half of the inverted U-shaped relation of EKC appears, the decrease of TCDE and CD may lead to the decrease of CCD.

The results also highlighted regional differences within the growth of low-carbon tourism in the YRB. The development extent of low-carbon tourism is thought to be directly correlated with regional differences in development [21]. There are significant differences in regional development among the nine provincial-level administrative regions in YRB. Taking 2019 as an example, Shandong's GDP is about 24 times that of Qinghai. It follows that there are clear regional variations in the growth of low-carbon tourism in YRB, which is not a finding of great importance. A more critical question is how regional development affects low-carbon tourism development. Tong et al. found that China's tourism economy has a significant carbon emission reduction effect, and although the direct effect of the tourism economy on carbon emission intensity is significantly positive, the indirect effect is significantly negative and stronger than the direct effect [59]. The findings of this paper suggest both the coordinated development of low-carbon TE and the total factor productivity of low-carbon tourism are related to regional differences. In addition, it is certain that regional differences in technical progress of low-carbon tourism lead to regional differences in total factor productivity of low-carbon tourism, and further cause regional differences in the development level of low-carbon tourism.

The study found that the total factor productivity of low-carbon tourism continued to increase, indicating that low-carbon tourism in YRB continued to be in a good development state from 2000 to 2019. The research results of Zhao et al. showed that the dependency of economic growth on fundamental energy usage is progressively diminished in YRB [60]. This study further proves that this conclusion is still valid in the tourism industry of YRB. Long-term carbon emissions are greatly reduced by renewable energy use whereas they increase substantially by using energy that is not renewable [61]. Therefore, reducing the energy consumption intensity of TE or increasing the proportion of renewable energy are two effective ways to improve the quality of low-carbon tourism development.

COVID-19 has certainly disrupted the tourism industry. It is difficult to effectively measure the carbon emissions of the tourism industry after the outbreak of COVID-19 due to the serious lack of data and the failure of some measurement coefficients. However, judging from the data on energy consumption in tourism-related industries, it is likely that technological advances in low-carbon tourism were not significantly affected by the outbreak. A more electricity-dependent tourism industry would produce fewer direct carbon emissions. For China, using more clean energy rather than coal to generate electricity is the key to achieving the carbon peaking and carbon neutrality goals.

As a study of low-carbon tourism assessment, this paper has the following novelties. (1) The assessment of low-carbon tourism development takes into account both the quantitative relationship between TCDE and TE, as well as their total factor productivity, which has been rarely seen in previous studies. (2) An improved CCD model is constructed, which can reflect the interaction between TCDE and TE. Previous studies related to low-carbon tourism tend to focus on decoupling analysis rather than coupling coordination analysis. However, in the early stages of development characterized by a rapid increase in carbon dioxide emissions, coupling coordination analysis may find more information. (3) The total factor productivity of low-carbon tourism is calculated to assess the quality of low-carbon tourism development. Total factor productivity is a mature assessment method of industrial development quality, but few studies have applied it to low-carbon tourism assessment. The total factor productivity of low-carbon tourism is not only related to TE and TCDE, but also related to the investment of tourism resources. With more and more attention paid to low-carbon tourism research, total factor productivity of low-carbon tourism may become one of the main methods to assess the development quality of low-carbon tourism.

# 5 Conclusions

Based on EKC, CCD model and MLPI, this study comprehensively assessed the development level of low-carbon tourism in YRB. Firstly, the carbon dioxide emissions related to tourism were estimated, and the TCDE index and TE index were calculated referring on the TCDE and TE indicators. Based on the results of the above indexes, it was judged whether the development process of low-carbon tourism in YRB had experienced EKC. Secondly, the coupling and coordination analysis of TCDE and TE was carried out to assess the interaction between the two systems. Finally, the total factor productivity of low-carbon tourism was assessed using MLPI, and the decomposition indices MLTE and MLTC were used to explore the impact of technical efficiency and technical progress.

The main findings and contributions of the paper were as followed: (1) The development of low-carbon tourism in YRB shows the characteristics of the initial EKC. TCDE and TE are growing, but the growth rate of TE is greater than the growth rate of TCDE. (2) There are substantial disparities in the CD of various regions in YRB, but the overall level is at a high level. The growth of TCDE and TE shows that they promote each other based on strong interaction. (3) Total factor productivity of low-carbon tourism has grown at an average yearly rate of 7% in YRB, primarily as a result of technical progress. The technical efficiency of all regions in YRB only fluctuates slightly, while technical progress has increased significantly in most regions. In addition, despite the impact of COVID-19 on the tourism economy in the Yellow River Basin, the pursuit of low-carbon tourism development remains ongoing. Overall, the quality of low-carbon tourism development in the YRB has been continuously improved, but to reduce TCDE, a long way still needs to be gone.

Although this study has achieved good results, there are still some limitations that cannot be avoided in future more detailed studies. First, due to the limitation of sample size and workload, the discovery of regional development differences in low-carbon tourism comes from the observation and summary of the basic characteristics of data results, and does not use more accurate methods such as spatial econometric analysis. Second, it is difficult to compare the development level of low-carbon tourism in different years or regions due to the lack of a comprehensive indicator. In future research, it is necessary to develop a comprehensive index that combines the assessment results of multiple aspects. Finally, this study did not consider the data during the COVID-19 pandemic. Investigating the impacts of the COVID-19 pandemic on the low-carbon tourism is an unavoidable issue for future research.

**Statements and Declarations**

**Data availability:** The dataset used in this study is available from the corresponding author upon request.

**Competing interests:** The authors have no relevant financial or non-financial interests to disclose.

**Author contributions: Xiaopeng Si:** Methodology, Software, Data Curation, Writing - Original Draft, Visualization. **Zi Tang:** Conceptualization, Methodology, Resources, Writing-Review & Editing, Funding acquisition.

**Acknowledgments**

This work was supported by the National Natural Science Foundation of China [grant number 41801137]; the National Social Science Foundation of China [grant number 22BJY157]; Projects of Philosophy and Social Sciences of Heilongjiang Province [grant number 20JYE275]; and Harbin University of Commerce Teachers "Innovation" Project Support Program in 2022 [grant number

22GLH075].